# Bayesian Network Enhanced with Structural Reliability Methods: Application

Daniel Straub[1] & Armen Der Kiureghian[2]

## Abstract

The enhanced Bayesian network (eBN) methodology described in the companion paper facilitates the assessment of reliability and risk of engineering systems when information about the system evolves in time. We present the application of the eBN (a) to the assessment of the life-cycle reliability of a structural system, (b) to the optimization of a decision on performing measurements in that structural system, and (c) to the risk assessment of an infrastructure system subject to natural hazards and deterioration of constituent structures. In all applications, observations of system performances or the hazards are made at various points in time and the eBN efficiently includes these observations in the analysis to provide an updated probabilistic model of the system at all times.

[1] Associate Professor, Engineering Risk Analysis Group, Technische Universität München, Arcisstr. 21, 80290 München, Germany. Email: straub@tum.de

[2] Taisei Professor of Civil Engineering, Dept. of Civil & Environmental Engineering, Univ. of California, Berkeley, CA 94720. Email: adk@ce.berkeley.edu

# 1 Introduction

In the companion paper (Straub and Der Kiureghian, 2010), we introduced the enhanced Bayesian network (eBN) methodology to combine Bayesian networks (BNs) with structural reliability methods (SRMs). The eBN facilitates the assessment of reliability and risk of engineering systems, in particular when the information about the system evolves in time. This is especially relevant in a near-real-time decision-support system, which enables the identification of optimal actions in large and complex systems immediately after new information becomes available. Examples of such applications include natural hazard warning systems, optimization of post-hazard mitigation actions, and structural health monitoring and management.

In this paper, we present the application of the eBN methodology to two example engineering systems, a structural system and an infrastructure system composed of structural systems. These are modified and extended versions of applications described in conference articles (Straub and Der Kiureghian, 2008 and 2009). In both examples, information on system elements and system performance, together with information on the hazards, become available over time, and are used to update the probabilistic models. The first example also demonstrates the facility of the proposed BN method for decision-making. Particular emphasis is placed on accurate modeling of the statistical dependencies in the system models, which, as will be demonstrated, are especially relevant in the context of information updating.

# 2 Reliability analysis and decision optimization of a structural system subject to environmental loading

## 2.1 Basic model of the structural system

Consider the structural system shown in Figure 1. This one-bay elasto-plastic frame under vertical load $V$ and horizontal load $H$ has been the subject of investigation by a number of authors following (Madsen *et al.*, 1986). Here, the model is according to (Der Kiureghian, 2005). The structural system is defined by the plastic-moment capacities $R_1, R_2, \ldots, R_5$ and the limit state functions corresponding to the three failure modes shown in Figure 1 are:

$$g_1(\mathbf{x}) = r_1 + r_2 + r_4 + r_5 - 5h$$
$$g_2(\mathbf{x}) = r_2 + 2r_3 + r_4 - 5v \quad (1)$$
$$g_3(\mathbf{x}) = r_1 + 2r_3 + 2r_4 + r_5 - 5h - 5v$$

The event of structural failure is defined through the failure domain:

$$\Omega_F(\mathbf{x}) = \left\{ \min_{i=1,2,3} g_i(\mathbf{x}) \leq 0 \right\} \quad (2)$$

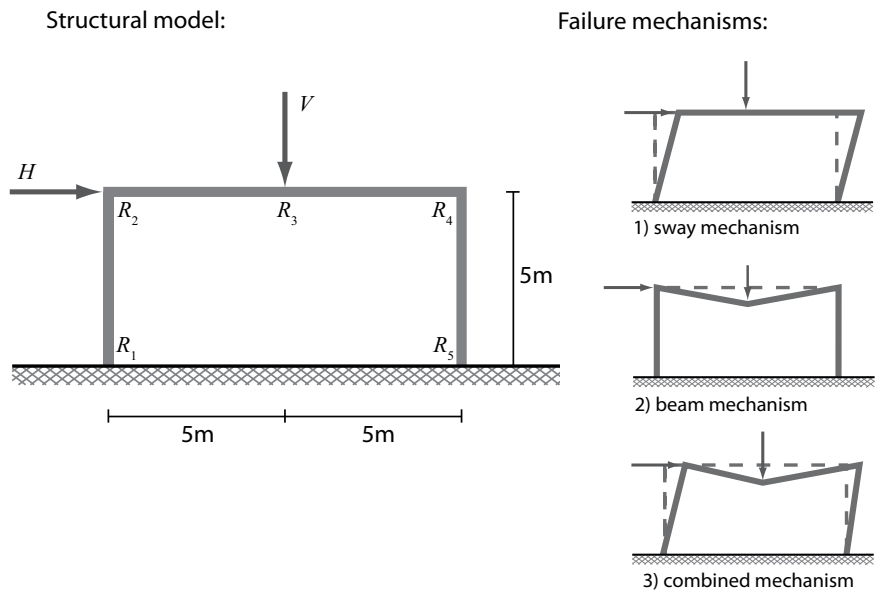

Figure 1. Example ductile frame and its failure mechanisms.

In the basic version of the problem, the plastic moment capacities $R_1, R_2, \ldots, R_5$ are equi-correlated Lognormal distributed random variables, with logarithmic mean $\lambda_R$, logarithmic standard deviation $\zeta_R$ and correlation coefficient $\rho_{\ln R} = \text{cov}(\ln R_i, \ln R_j) / \zeta_R^2, i \neq j$. The gravity load $V$ and the environmental load $H$ are statistically independent random variables. The full probabilistic model for the basic case of this example is summarized in Table 1.

*Table 1. Probabilistic model of example structural system (Der Kiureghian, 2005).*

| Variable | Distribution | Mean | c.o.v. | Correlation |
|---|---|---|---|---|
| $R_i$, i=1,...,5 [kNm] | Joint Lognormal | 150 | 0.2 | $\rho_{lnR} = 0.3$ |
| $H$ [kN] | Gumbel | 50 | 0.4 | Independent |
| $V$ [kN] | Gamma | 60 | 0.2 | Independent |

As a first step, we consider the classical reliability problem of determining $\Pr(F) = \Pr(\mathbf{X} \in \Omega_F)$. In Figure 2, the corresponding eBN is shown. The variable $E$ of this model represents the state of the system, i.e., either failure $\{E = F\}$ or survival $\{E = \bar{F}\}$ occurs. In this formulation, we make use of the fact that the equi-correlation among the random variables $R_1, R_2, \ldots, R_5$ can be modelled through a common parent variable $U_R$ with standard Normal distribution, such that $R_1, R_2, \ldots, R_5$ are conditionally independent given $U_R = u_R$, with conditional distributions

$$F(r_i | u_R) = \Phi\left(\frac{\ln r_i - u_R \sqrt{\rho_{lnR}} - \lambda_R}{\sqrt{\zeta_R^2 - \rho}}\right), \quad i = 1, \ldots, 5 \tag{3}$$

where $\Phi$ is the standard Normal cumulative distribution function (CDF). More generally, statistical dependence can be represented in this way among groups of random variables whose correlation matrix is of the Dunnet-Sobel class (Dunnett and Sobel, 1955).

As observed in Figure 2, the only discrete variable in the eBN is the system state $E$. Therefore, reducing this network to the corresponding reduced BN (rBN), which consists only of the single variable $E$, is identical to directly solving the structural reliability problem for the probability of event $F$. Except for the graphical illustration of the model, nothing is gained from the eBN approach in this basic example.

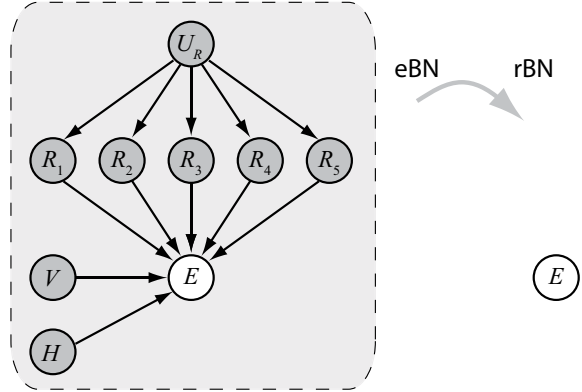

*Figure 2. The eBN model and the corresponding reduced rBN for the unconditional reliability problem. The shaded area is the Markov envelope.*

## 2.2 Reliability analysis with measurements of member capacities

Since the eBN is powerful for Bayesian updating, we consider a case where information becomes available through in-situ, non-destructive measurements of the plastic moment capacities at locations 4 and 5. The measurements are indirect and associated with error, such that the resulting measurements are

$$M_i = R_i + \varepsilon_i, \quad i = 4,5 \qquad (4)$$

with the $\varepsilon_i$ being Normal distributed with mean 0 and standard deviation 15 kNm. In the eBN model, the variables $M_i$ are represented by discrete variables, since we expect to obtain evidence on these variables. A first (naïve) modelling approach is to simply discretize the variables $M_4, M_5$ and to construct the corresponding eBN as shown in Figure 3. The discretization follows the procedure described in the companion paper. The resulting rBN is presented on the right-hand side of Figure 3.

The rBN shown in Figure 3 is not reflecting causality, since the measurements cannot influence the capacities (only our estimates of the capacities). This corresponds to a diagnostic model (Jensen, 2001). As discussed in the companion paper, causality is preferred, because it facilitates understanding and communication of the model, but also because it often leads to simpler calculations. This is the case here, as we show by introducing an alternative model. To maintain causality, the measured variables $R_4$ and $R_5$ must be included in the rBN. To this end, an alternative eBN model is established, wherein these variables are discretized in accordance to the method described in Section 4 of the companion paper, as shown in Figure 4. Note that the resulting rBN maintains causality between the measured variables and the measurement outcomes.

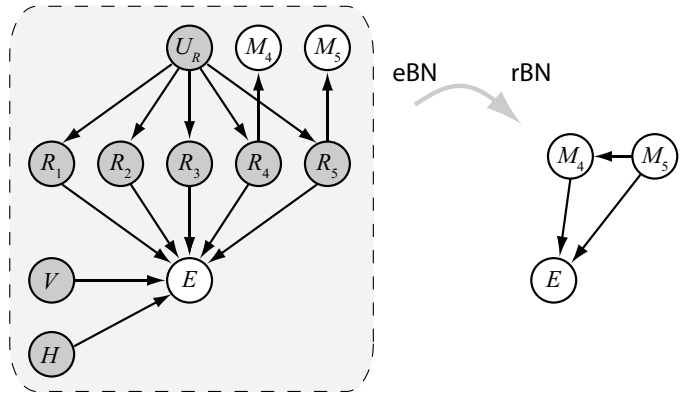

*Figure 3. The eBN model and the corresponding rBN for the problem involving measurements of element capacities.*

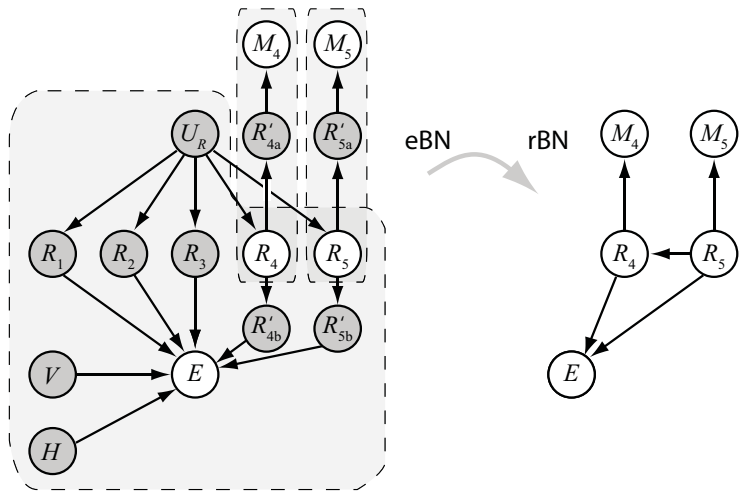

*Figure 4. The alternative eBN model, with corresponding causal rBN, for the problem involving measurements of element capacities.*

As observed from the respective Markov envelopes, the first model (Figure 3) requires computation of $\Pr(F\,|\,M_4 = m_4 \cap M_5 = m_5)$ through SRM, whereas the second model (Figure 4) requires computation of $\Pr(F\,|\,R_4 = r_4 \cap R_5 = r_5)$ through SRM. The latter

computation corresponds to performing the unconditioned system reliability analysis whereby the random variables $R_4$ and $R_5$ are replaced with deterministic values $r_4$ and $r_5$, which is simpler than directly solving for the conditional probability $\Pr(F | M_4 = m_4 \cap M_5 = m_5)$, and might result in more accurate probability estimates. This model is utilized in the following. Table 2 summarizes the results obtained with the eBN-rBN model, compared with results from Monte Carlo simulation (MCS) for verification. The reliability index in this table is computed as the standard normal variate for the complement of the failure probability. The conditional MCS results are obtained with the method proposed in (Straub 2011), which enables the computation of reliability conditional on measurements by means of simulation techniques. The eBN-rBN results are in agreement with the MCS results, as evident from Table 2.

*Table 2. Reliability index (failure probability) conditional on measurement results, $\Pr(F | M_4 = m_4 \cap M_5 = m_5)$. For Monte Carlo simulation (MCS), the 95% probability interval is presented.*

| Measurements [kNm]: | No measurement | $M_4$=50, $M_5$=100 | $M_4$=150, $M_5$=100 | $M_4$=150, $M_5$=200 |
|---|---|---|---|---|
| eBN-rBN Model: | 1.94 (2.6·10$^{-2}$) | 0.70 (0.24) | 1.80 (3.6·10$^{-2}$) | 2.45 (0.71·10$^{-2}$) |
| MCS (10$^6$ samples): | 1.93–1.94 | 0.63–0.77 | 1.78–1.81 | 2.44–2.48 |

The above results are computed with discretization of $R_4$ and $R_5$ in 21 states with interval borders (50:10:250)kNm. Therefore, to establish the rBN, $21^2 = 441$ SRM computations are required in the construction of the conditional probability mass matrix (PMF) of $E$. These are carried out by the First Order Reliability Method (FORM). Since these are straightforward series system reliability problems, the SRM computations are performed efficiently in an automated manner. The computer program CalREL (Der Kiureghian *et al.*, 2006) is used for this purpose. The 21 values of the conditional PMF of each $M_i$ given $R_i=r_i$ are obtained by use of Eq. (4) and the normal CDF. Once the rBN is established, the updating computations shown in Table 2 take negligible time.

## 2.3 Life-cycle reliability analysis with measurement of capacities and observations of past performances

Let the environmental load *H* acting on the structure shown in Figure 1 vary with time. Since only the extreme load is of interest, we consider this temporal variability through the annual maximum load $H(t)$ in any year $t$, $t = 1, 2, \ldots, 20$, during the anticipated 20 years service life of the structure. Although it is common to assume that the $H(t)$ for different years are statistically independent of each other, this is not generally the case due to common influencing factors, such as model and statistical uncertainties, see, e.g., (Coles and Simiu, 2003). Modeling of such common factors is straightforward in the eBN approach. One simply formulates the $H(t)$ conditional on the common influencing factors. To exemplify the accounting of this effect, we expand the model presented earlier by defining the parameter $u_H$ of the distribution of $H(t)$, $F_{H(t)}(h) = \exp\{-\exp[-\alpha_H(h - u_H)]\}$, as random variable $U_H$. The gravitational load *V* is assumed to have only one realization during the service life of the structure. The corresponding eBN is presented in Figure 5, including the measurements $M_4$ and $M_5$ considered in the previous example. The links from $E(t)$ to $E(t+1)$ model the fact that a structure that has failed at time $t$ remains in the failed state (no replacement or repair is assumed). The node $E(t)$ thus represents the state of the structure up to time $t$.

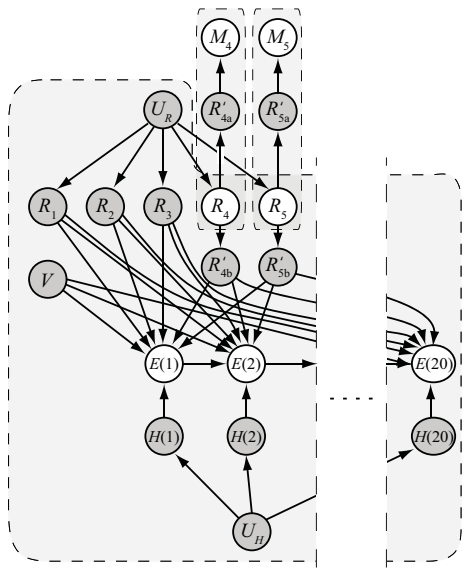

*Figure 5. A naïve eBN model for the life-cycle reliability analysis.*

The eBN in Figure 5 is computationally inefficient. Because all performance variables $E(1), \ldots, E(20)$ are part of the same Markov envelope, establishing the corresponding rBN would require computation of the joint distribution of these variables. Furthermore, one variable $E(t)$ would have all other performance variables as parents in the rBN. (See the case illustrated in Figure 7a of the companion paper.) To improve the model, we consider the modeling strategies presented in the companion paper. First, we note that, by discretizing the parent nodes of $E(1), \ldots, E(20)$, the latter can be modeled as statistically independent given their parent nodes. However, as observed from Figure 5, this would require discretization of $R_1, \ldots, R_5$, $U_H$ and $V$, and SRM computations would be required for all combinations of the states of these discretized variables. To avoid this, we use the divorcing strategy presented in Section 4 of the companion paper. For this purpose, we introduce a continuous variable $Q$ in the model, which represents the capacity of the structural system with respect to the environmental loads $H(t)$. By discretizing this variable we achieve an efficient model of the structural system over its entire service life, as shown in Figure 6. In this model, $U_H$ and the $H(t)$'s are discretized as well. While this is not required for computational efficiency, it is included in the model to enable updating when observations of past loads become available.

The variable $Q$ is discretized over a set of intervals. Let $q_k^+$ denote the upper limit of the $k$-th interval. The CDF of the discretized variable $Q$ is then obtained from the CDF of the corresponding continuous random variable at the upper limit of each interval. The latter is defined through the domain $\Omega_Q^{(k)}(\mathbf{x})$ as

$$\Pr(Q \leq q_k^+) = \int_{\mathbf{x} \in \Omega_Q^{(k)}(\mathbf{x})} f(\mathbf{x}) d\mathbf{x} \tag{5}$$

The domain $\Omega_Q^{(k)}(\mathbf{x})$ is obtained by replacing $h$ in Equation (1) with $q_k^+$:

$$\begin{aligned} g_{Q1}(\mathbf{x}, q_k^+) &= r_1 + r_2 + r_4 + r_5 - 5q_k^+ \\ g_{Q2}(\mathbf{x}, q_k^+) &= r_2 + 2r_3 + r_4 - 5v \\ g_{Q3}(\mathbf{x}, q_k^+) &= r_1 + 2r_3 + 2r_4 + r_5 - 5q_k^+ - 5v \end{aligned} \tag{6}$$

and

$$\Omega_Q^{(k)}(\mathbf{x}) = \left\{ \min_{i=1,2,3} g_{Qi}(\mathbf{x}, q_k^+) \leq 0 \right\} \quad (7)$$

The limit state function and the corresponding domain representing the failure of the structural system at time *t* then become

$$g_{F(t)}[q, h(t)] = q - h(t) \quad (8)$$

$$\Omega_{E(t)}^{(0)}[q, h(t)] = \left\{ g_{F(t)}[q, h(t)] \leq 0 \right\} \quad (9)$$

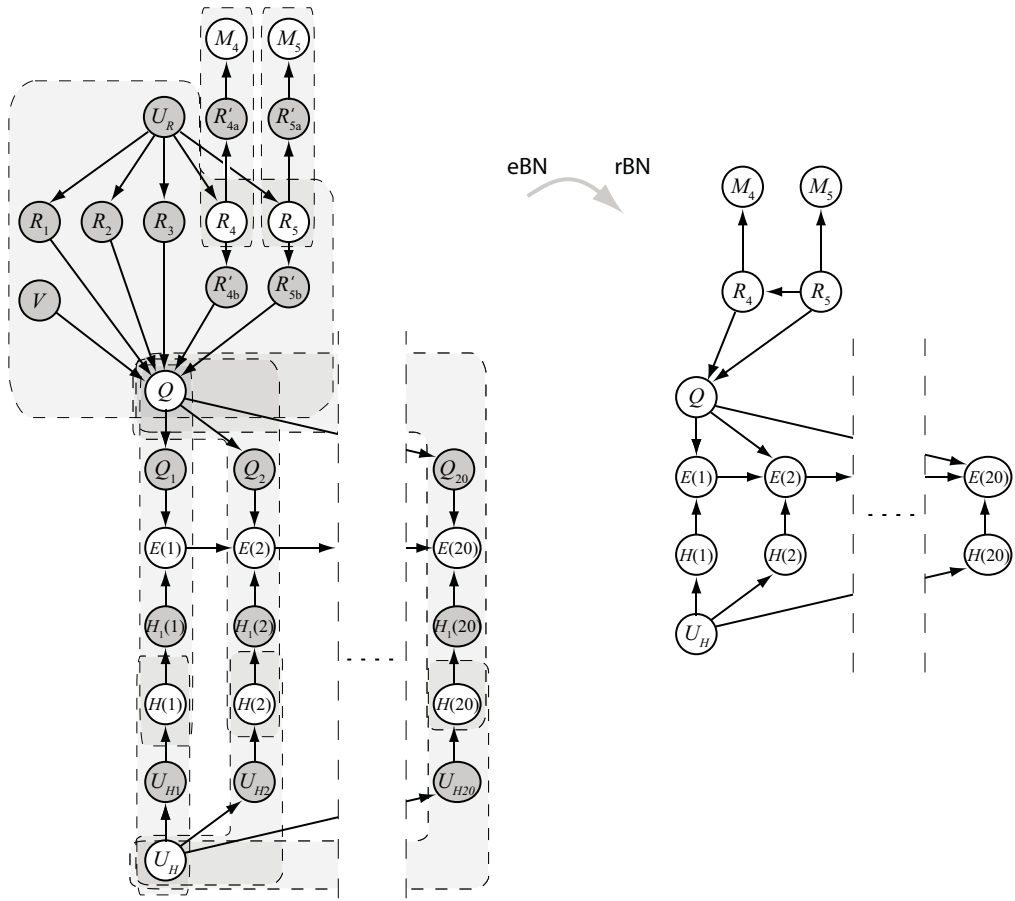

*Figure 6. An efficient eBN model for the life-cycle reliability analysis, with the corresponding rBN.*

For numerical evaluation, the same probabilistic model as earlier is employed, with the exception of modified distribution parameters for $H(t)$: $\alpha_H = 0.0641\text{kN}^{-1}$ and $U_H$ a Lognormal random variable with mean 35kN and standard deviation 10kN. (In the first example, $\alpha_H$ was the same and $u_H$=41.1kN.) The probabilistic model is summarized in **Error! Not a valid bookmark self-reference.**.

*Table 3. Probabilistic model of the structure with time-varying environmental load H(t).*

| Variable | Distribution | Mean | c.o.v. | Correlation |
|---|---|---|---|---|
| $R_i$, $i$=1,...,5 [kNm] | Joint Lognormal | 150 | 0.2 | $\rho_{\ln R} = 0.3$ |
| $\varepsilon_i$, $i$=4,5 [kNm] | Normal | 0 | 15 | Independent |
| $V$ [kN] | Gamma | 60 | 0.2 | Independent |
| $U_H$ [kN] | Lognormal | 35 | 0.286 | Independent |
| $H(t)$, $t$=1,...,20 [kN] | Gumbel | $u_H$+9.0 | 20/($u_H$+9.0) | Independent |

In Figure 7 we plot the reliability index of the structure as a function of time for different evidence cases. If the structure has survived the first five years, the reliability over the remaining years increases because (a) the probability of failure for the first five years is zero, and (b) the fact that the structure has withstood the loads in the first five years leads to a modified posterior distribution of the capacity (a proof-loading effect). In particular, the beam-failure mechanism, which only depends on the time-invariant vertical load, will not occur given that the structure has survived the first year. In the case where the environmental load in year 5 is additionally observed to be 80kN, the updated reliability is slightly lower than that without the observation of the load. This result is surprising at first sight, because the observation of survival to such a large load is expected to increase the proof-loading effect. However, because the loads $H(t)$ are correlated due to the common model uncertainty in $U_H$, observation of a large $H(5)$ increases the posterior probability of large environmental loads in future years. Accordingly, if additional observations of low environmental loads, $H(t)$=30kN, are assumed to have been made in years $t$=1,...,4, the resulting reliability indexes increase. To further demonstrate this effect, the posterior distribution of $U_H$ is plotted in Figure 8 for these cases, showing a shift towards large values of $u_H$ for the observation in year 5 alone and a shift towards smaller values of $u_H$ for the observations in all 5 years.

In Figure 9, observations of performance are considered together with capacity measurements. It is observed that the latter have a significant influence on the reliability index.

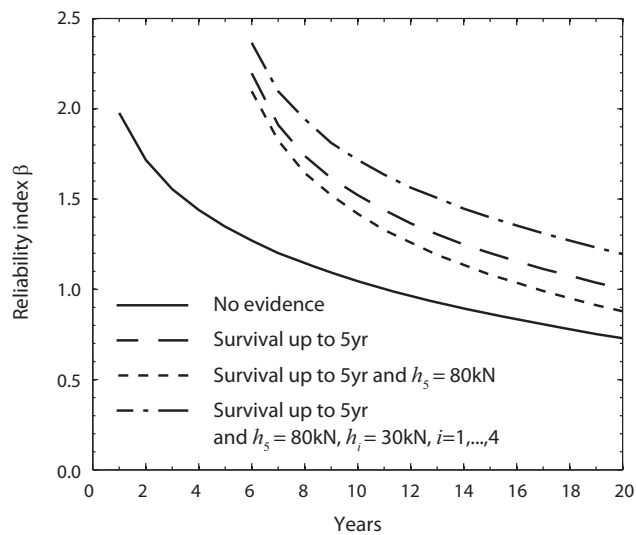

Figure 7. Reliability indexes computed for different evidence cases.

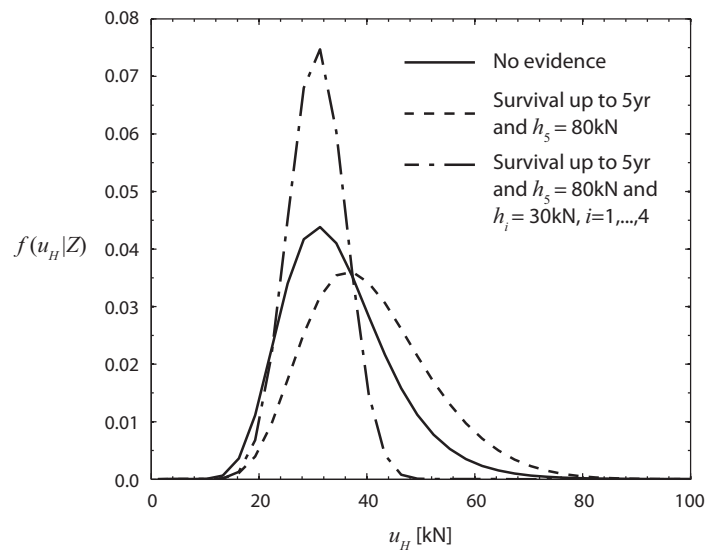

Figure 8. Posterior PDF of $U_H$ conditional on different evidence cases.

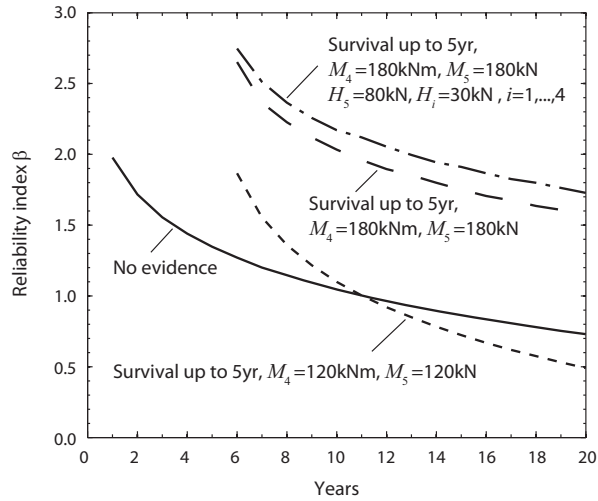

*Figure 9. Reliability indexes computed for different evidence cases that include measurements of element capacities.*

The above results are computed with discretization of $R_4$ and $R_5$ in 21 states with interval borders (50:10:250)kNm and discretization of $Q$ in 31 states with interval borders (0:5:150,∞)kN. Therefore, to establish the rBN, $21^2 \times 31 = 19{,}251$ SRM computations are required. Using FORM, these computations require approximately one CPU hour on a standard Pentium II computer. It is noted that the number of discrete states can likely be reduced without significant loss of accuracy, but no optimization was considered here, since, as in the first example, the employed FORM algorithm had no problem with the straightforward system reliability analyses. The conditional PMF of $E$ given $Q=q$ and $H_t=h_t$ is obtained through numerical integration, as is the conditional PMF of $H_t$ given $U_H=u_H$. The discretization scheme of $H_t$ is identical to that of $Q$ with interval borders (0:5:150,∞)kN and the one for $U_H$ is (0:3:150,∞)kN. Once the rBN is established, computation of any of the curves shown in Figures 7-9 takes in the order of one CPU second on a standard Pentium II computer.

## 2.4 Value of information analysis for capacity measurements

This section illustrates the application of the eBN model to decision optimization. It makes use of the facility to extend any BN to a decision graph (aka influence diagram) by including decision and utility nodes (Shachter, 1986; Jensen, 2001). No introduction to the theory of decision graphs is given here, but the provided description of its application is self-contained.

We consider two decisions. The first is on the optimization of mitigation actions. Upon knowing the outcomes of the measurements, the owner of the structure, i.e., the decision-maker, may decide to replace the elements of the structure if the measurement outcomes are unfavorable. The eBN can support the decision-maker in finding the optimal alternative. The second decision that is considered is on the measurements. Thus far we have assumed that measurements will be performed. However, at an earlier stage it has to be decided if and what to measure, depending on the cost of measurement. We employ the eBN extended to a decision graph to determine the benefit of the measurements, the so-called *value-of-information*, and whether the measurements should be performed in the first place. The first type of decision analysis is known as terminal analysis, the second as preposterior analysis (Raiffa and Schlaifer, 1961).

The decision and utility nodes are included directly in the rBN, as shown in Figure 10. This graph includes a decision node (rectangular) for the replacement/no replacement decision; and two utility nodes (diamond-shaped), representing the cost of the replacement and the cost of failure of the structure. (This model assumes that the time of failure of the structure is immaterial. A more realistic model would have the utility node as a function of the failure events at different years to enable inclusion of discounting.)

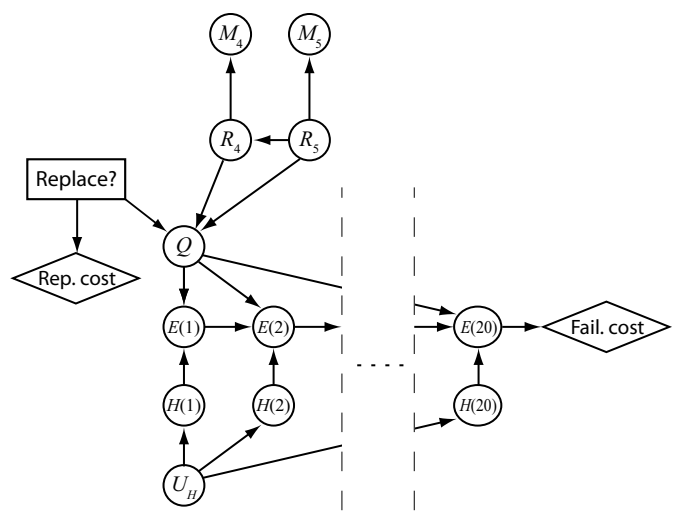

*Figure 10. Decision graph based on the rBN in Figure 6.*

A simple replacement model is selected: If the elements of the structure are replaced upon an unfavorable measurement, then the new structure is represented by the same stochastic model as the original structure. (In case of an unfavorable measurement, the decision-maker considers that a new draw from the same population is likely to produce a more favorable outcome.). In the rBN, the replacement model is represented by the PMF of $Q$, which must now be defined conditional on the replacement action. Given "no replacement" the PMF is identical to the one applied previously. Given "replacement", the PMF of $Q$ becomes statistically independent of $R_4$ and $R_5$, $p(q|r_4,r_5,a=\text{'replacement'}) = p(q|a=\text{'replacement'})$, and corresponds to the PMF of $Q$ for the original structure when no evidence is available.

The utility nodes describe the consequences for given values of their parents. The utility node for the replacement cost takes the value $-10^4$ for the case of replacement and the value zero otherwise. The utility node for the failure cost takes the value $-10^5$ for the case of failure and the value zero otherwise.

The decision optimization proceeds by computing the expected utility for the different decision alternatives. Many of the available BN inference software perform these calculations; otherwise, the rBN provides the conditional PMF's required for the evaluation of the expected utility. For the first decision problem, the expected utilities for the two decision alternatives ("replace" and "do not replace") are shown in Table 4 for several evidence cases.

*Table 4. Expected utilities for the decision alternatives "replace" and "do not replace" for different evidence cases. Bold numbers indicate optimal decisions.*

| Evidence | Expected utility for "do not replace" | Expected utility for "replace" |
| --- | --- | --- |
| a) No evidence | **–23,276** | –23'276 + –10'000 = –33,276 |
| b) $m_4$ = 120kNm, $m_5$ = 120kNm | –44,087 | –23'276 + –10'000 = **–33,276** |
| c) $m_4$ = 180kNm, $m_5$ = 180kNm | **–7,936** | –23'276 + –10'000 = –33,276 |
| d) $m_4$ = 100kNm, survival up to $t$=5yr | –32,191 | –19'421 + –10'000 = **–29,421** |
| e) $m_4$ = 100kNm, survival up to $t$=10yr | **–20,814** | –14'798 + –10'000 = –24,798 |

As expected, nothing is gained from replacement in the no-evidence case (a), since in this case the new structure is described by the same stochastic model as the original one. In case of the unfavorable measurement outcomes (b), the replace decision is the optimal one. In case of the favorable outcomes (c), no replacement should be performed. In case (d), where an unfavorable measurement result is obtained in year 5, the structure should be replaced, despite the fact that it did not fail up to this time. In case (e), where the same measurement result is obtained in year 10, the optimal decision is not to replace. In this case, the proof-loading effect and the fact that the remaining service life is only 10 years are in favor of the "do not replace" option.

To determine whether either measurement should be conducted, we perform a value-of-information (VOI) analysis. In this analysis, the rBN is used to compute the optimal terminal decision and the associated expected utility conditional on all possible outcomes of each considered measurement, $\mathrm{E}[U\,|\,M_i = j]$. To obtain the expected utility prior to knowing the measurement outcome, but given that the measurement will be performed, the expectation with respect to each $M_i$ is computed:

$$\mathrm{E}[U\,|\,a_i] = \sum_j \mathrm{E}[U\,|\,M_i = j]\Pr[M_i = j] \qquad (10)$$

where $a_i$ denotes the decision to perform measurement $i$ and $\Pr[M_i = j]$ is the prior probability of $M_i$ being in state $j$. In Equation (10), the cost of the measurement itself is not included. The VOI of $a_i$ is computed as the difference between the expected utility given $a_i$ and the expected utility when no measurement is conducted:

$$VOI(a_i) = \mathrm{E}[U\,|\,a_i] - \mathrm{E}[U] \qquad (11)$$

Some BN inference software perform these calculations directly, e.g., the free GeNIe BN modeling environment developed by the Decision Systems Laboratory of the University of Pittsburgh.

For the present example, the results for the individual measurements are $VOI(a_4) = 1,802$ and $VOI(a_5) = 1,168$ of utility units. This shows that more information is gained by measuring capacity $R_4$, since this variable has a higher importance, as is evident from the limit state functions defining the system failure, Eq. (1). (Note that $r_4$ appears in all three limit-state functions and it has a coefficient of 2 in $g_3$, while $r_5$ appears only in $g_1$ and $g_3$, both with coefficients 1.) The two measurements are not independent, and to determine whether both should be performed, it is necessary to also compute the joint *VOI* of the two measurements. Here, we obtain $VOI(a_4, a_5) = 2,763$ units. If the cost of a measurement is lower than its *VOI*, it is beneficial to perform it. As an example, if one measurement costs 500 units, then both measurements should be performed. If one measurement costs 1,500 units of utility, then only measurement $a_4$ should be performed, since its VOI is 1,802 and the VOI of both measurements is only 961 ($= 2763 - 1802$) units higher. Therefore, the second measurement would not be cost effective in that case.

## 3 Reliability analysis of an infrastructure system subject to environmental loading and deterioration

This example demonstrates the relevance of the eBN methodology for reliability and risk assessment of infrastructure systems, which are comprised of structural and non-structural systems. Examples of structural systems that are part of infrastructure systems are bridges and tunnels that are elements of transportation systems, and offshore platforms and pipelines that are elements of an oil and gas production system; examples of non-structural elements in these systems are power supply units, pumping stations or traffic control systems. In principle, the eBN might also include models of the related organizational and behavioral aspects, such as the reaction of personnel or the public to extreme events, but this is not considered here.

Assessment of infrastructure system performance and risk must account for both the temporal and spatial dimensions of the problem. In the temporal dimension, the analysis must include accounting for:

- temporal statistical dependence of hazards (as considered in the previous example);
- deterioration of structural and non-structural elements;

- discounting.

In the spatial dimension, the analysis must include accounting for:

- spatial distribution of the hazard (demand);
- spatial distribution of the system capacities;
- network system functionality.

These temporal and spatial aspects determine the dependences among the system elements. eBNs and corresponding decision graphs are well suited for efficiently modeling these system dependences. However, the particular advantage of the eBNs is that they facilitate updating of the uncertainties in the system model based on observations at any scale of the system. For example, monitoring of a structural component in a bridge can be utilized to update the probabilistic model of the entire infrastructure. Such multi-scale system analysis is possible because the eBN framework enables inclusion of structural reliability methods in a general BN model of an infrastructure system.

## 3.1 Object-oriented BN

To facilitate establishing and representing the eBN and the corresponding rBN models for infrastructure systems, we make use of object-oriented BNs (OOBNs) (Koller and Pfeffer, 1997). In an OOBN, a class is a Bayesian network in which some of the variables are defined as input and some as output (the attributes of the class). The instantiations of the classes (the objects) are embedded in a higher level BN, with which they communicate through input and output variables. The OOBN methodology includes the general concepts of object-oriented programming, such as inheritance from a class to a subclass. However, for the purpose of the considered application, it is sufficient to think of the OOBN as a BN in which sets of variables are grouped into objects, which are either a part of other, higher-level objects or directly of the top-level model. To perform inference, the OOBN is treated like a large BN. This concept, which is quite intuitive, is illustrated in Figure 11, in which rounded rectangles represent objects. As an example, the typical "bridge" object is connected through its input variable "spectral acceleration," which is an output variable of the "earthquake characteristics" object, and its output variable "performance," which is an input variable to the "transportation system" object.

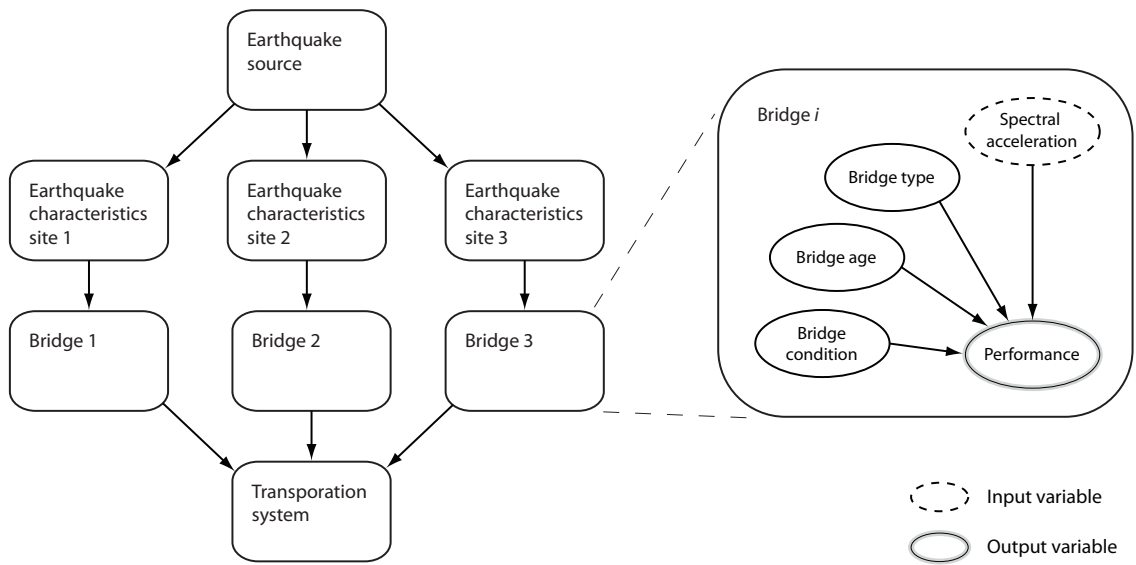

*Figure 11. Illustration of the OOBN concept, considering an example of a transportation network with three bridges subject to earthquakes.*

## 3.2 eBN-based framework for infrastructure risk analysis

A generic OOBN model framework for infrastructure systems is presented in Figure 12. The horizontal and vertical directions of the model respectively describe the temporal and spatial dimensions of the problem. All nodes are objects (thus their rounded rectangular shapes), i.e., they generally represent lower-level BNs. Each object can contain both continuous and discrete variables, in which case it is an eBN, or only discrete variables, in which case it is a BN.

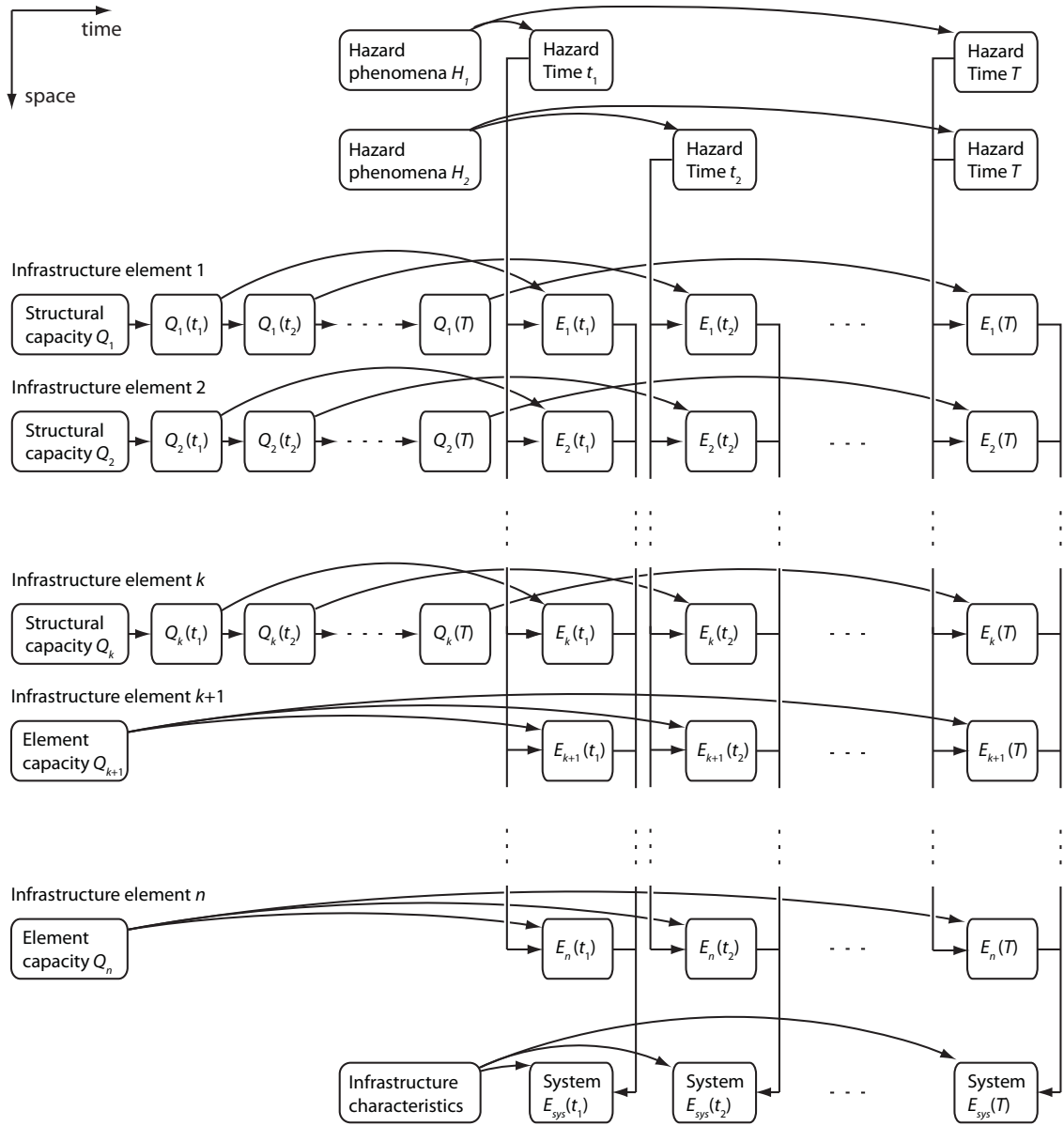

*Figure 12. A spatial-temporal eBN model framework for infrastructure systems.*

The framework includes objects that represent relevant natural hazard phenomena. These objects describe the general characteristics of the hazards for the considered region. They have as children instantiations of the hazard classes, each of which models a particular instance of a hazard at time $t_i$ (e.g., a particular windstorm or earthquake event) at the regional level. Next, the framework includes objects $Q_j$ that represent the time-invariant characteristics of the infrastructure elements, such as bridges and tunnels. If $Q_j$ is the capacity of a structural system, and if that system is subject to deterioration, then objects $Q_j(t_i)$ are introduced to model the capacity as a function of time. The performance of each system element $j$ at time $t_i$ is modeled by an object $E_j(t_i)$ that includes the local hazard characteristics, such as the spectral acceleration at the site due to an earthquake or the local wind speed due to a regional windstorm. Finally, the performance of the infrastructure system is modeled by the object $E_{sys}(t)$, which represents the system functionalities, e.g., the possibility of travel between certain origin-destination nodes.

The suggested framework makes certain assumptions about conditional independence among the elements of the model. For example, performances of different infrastructure elements are assumed to be statistically independent for given hazard and unknown infrastructure system performance, and deterioration of elements is modeled as a Markovian process. However, as shown in (Straub 2009) for the case of deterioration, these assumptions do not represent strict limitations on the model, since the independence assumption can be circumvented by adding additional variables. Therefore, the above framework is considered sufficiently general for most types of infrastructure systems and hazards. The approach, however, does have limitations from the viewpoint of computability. These stem from the need to limit the sizes of the Markov envelopes in the eBN and the requirement to limit the complexity of the resulting rBN. Below, we briefly describe strategies to deal with these computational limitations. The implications of these limitations on the applicability of the model for realistic infrastructure systems is discussed later, following the example.

First, regarding the sizes of the Markov envelopes, to the extent possible these should be confined to the individual objects of the general framework. A general modeling strategy to achieve this objective is to discretize any of the input and output variables of the framework objects that are of continuous type. The method described in Section 3.8 of the companion paper can be used for this purpose. Second, to limit the complexity of the resulting rBN, the number of input and output variables of each object in the general framework, as well as the number of states of these variables, should be made as small as possible. Here, the modeler must find a balance between accuracy and efficiency of the model. The modeling strategies introduced in the companion paper and demonstrated for the first example of this paper can be helpful in achieving this goal. Third, like any engineering model, an eBN model should be targeted towards specific decision problems.

In doing so, those parts of the general framework that are immaterial to the particular decision problem can be omitted. Furthermore, the number of states of the discrete variables, which determine the accuracy of the model, should be selected in correspondence with the accuracy of the available information and the relevance of the variable to the decision problem at hand.

## 3.3 Example infrastructure reliability and risk analysis

Consider the simplified train transportation system illustrated in Figure 13. The rectangles in this network represent structural elements (bridges), the squares represent non-structural elements (control systems), and the lines represent railways. At time $t$, each of the structural and non-structural elements in the system is either in the survival state, $E_j(t)=1$, in which case the corresponding link is connected, or in a failed state $E_j(t)=0$, in which case the corresponding link is disconnected. The railway segments are always in the working state. The performance of the system at any given time $t$ is defined as a binary variable: $E_{sys}(t) = 1$ if there is connectivity between cities A and B, and $E_{sys}(t) = 0$ otherwise.

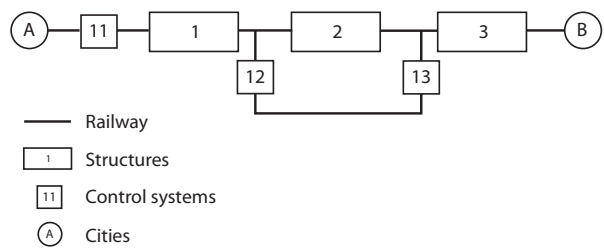

*Figure 13. Example transportation system.*

To simplify the presentation, we assume that all structural systems in our infrastructure are represented by the idealized structural model of the first example, as shown in Figure 1, and that the probabilistic description is as in Table 1. Inclusion of different models for different structures would be straightforward and would not increase the computational effort, but would require providing additional details. The system is subject to an environmental hazard (e.g., windstorm, earthquake), which at the level of structure $j$ is modeled by the variable $H_j(t)$ that is defined conditional on the regional hazard characteristics $H(t)$ as $H_j(t) = H(t) \cdot X_j(t)$. Here, $X_j(t)$ represents the effect of the local site conditions. In a real application, $X_j(t)$ would be replaced by functions that depend on the site location and hazard characteristics (e.g. attenuation models for the case of earthquakes), but conceptually the model remains the same, and so do the associated computational efforts. An exception is the case where $X_j(t)$ for different sites are statistically dependent, such as is the case for spatially correlated error terms in earthquake attenuation models (Straub *et al.*, 2008). Deterioration is introduced in the model through the conditional PMF of the capacity of the structure at time *t*, $\Pr[Q_j(t) = q_j(t) | Q_j(t-1) = q_j(t-1)]$. Any Markovian deterioration model can be represented in this form. For the present example we assume that the capacity at any time step *t* is Beta distributed in the interval $[0, q_j(t-1)]$, with mean value $0.98 \, q_j(t-1)$ and standard deviation $0.01 \, q_j(t-1)$.

All non-structural elements in the system are of the same type. For the element at site *j*, the failure probability conditional on the local hazard intensity, $H_j(t)$, is given through a Lognormal fragility function $\Pr[F_j(t) | h_j(t)] = \Phi\{[\ln h_j(t) - \lambda_j]/\zeta_j\}$ with parameters $\lambda_j$ and $\zeta_j$. The non-structural elements do not deteriorate and their capacities remain invariant of time.

The full probabilistic model of the infrastructure system is summarized in Table 5. The objects of the eBN model of the infrastructure are shown in Figure 14. The infrastructure performance is modeled using the "explicit connectivity" approach described in (Bensi *et al.*, 2009). Unlike the life-cycle example for the single structural system, the variables $E_j(t)$ here are defined as the performances of the infrastructure elements in year *t*, not the performances up to year *t*. This model is more appropriate if the interest is in the reliability of the infrastructure system, since it is realistic to assume that a failed system element would be immediately replaced. However, in order to account for implicit proof-loading effects, the reliability in year *t* must be computed conditional on survival up to year *t*−1.

*Table 5. Probabilistic model for the infrastructure example.*

| Variable | Distribution | Mean | c.o.v. |
|---|---|---|---|
| $R_{ji}$, $i=1,...,5$, $j=1,2,3$ [kNm] | Joint Lognormal | 150 | 0.2 |
| $\varepsilon_{ji}$, $i=4,5$, $j=1,2,3$ [kNm] | Normal | 0 | 15 |
| $V$ [kN] | Gamma | 60 | 0.2 |
| $U_H$ [kN] | Lognormal | 35 | 0.286 |
| $H(t)$, $t=1,...,10$ [kN] | Gumbel | $u_H+9.0$ | $20/(u_H+9.0)$ |
| $Q_j(t)$, $j=1,2,3$, $t=2,...,10$ | Beta in $[0, q_j(t-1)]$ | $0.98q_j(t-1)$ | 1/98 |
| $X_j(t)$, $j=1,2,3$, $t=1,...,10$ | Lognormal | 0.8 | 0.1 |
| $\lambda_j$, $j=11,12,13$ | Deterministic | 4.76 | |
| $\zeta_j$, $j=11,12,13$ | Deterministic | 0.246 | |

The logarithms of the moment capacities $R_{ij}$ and $R_{kj}$ within a structure $j$ are correlated with $\rho_{\ln R} = 0.3$.

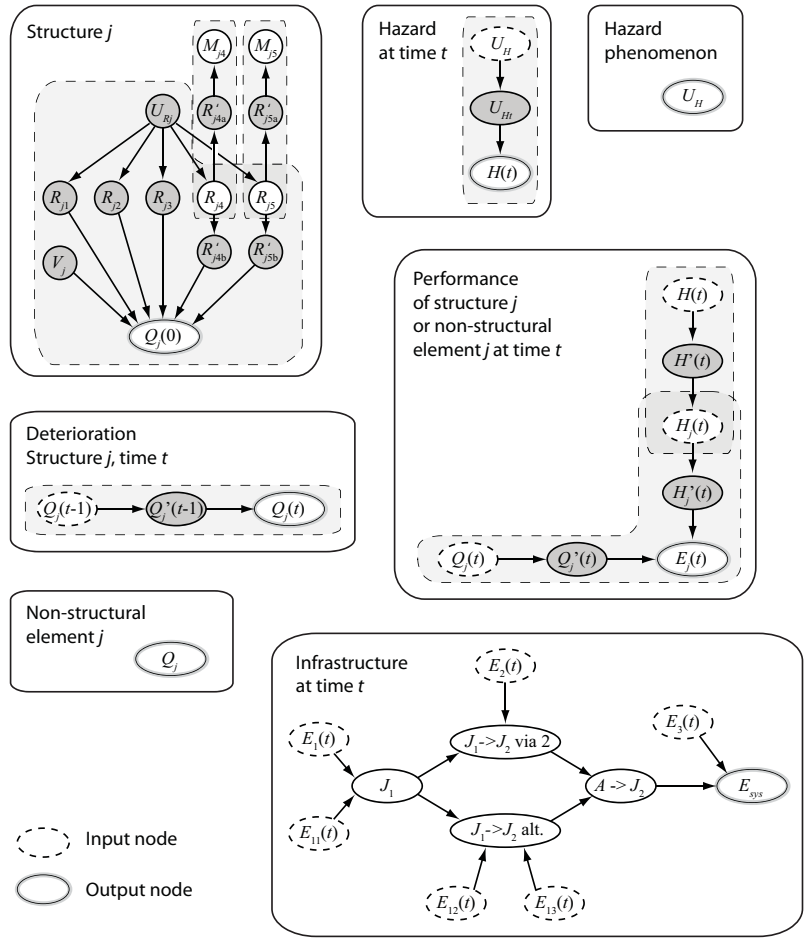

*Figure 14. eBN for the example infrastructure. The objects are connected through their input and output nodes in accordance with the general framework in Figure 12.*

To demonstrate the potential of the eBN framework as a tool for near-real-time infrastructure risk assessment, the model is applied to determine the reliability of the network as a function of time with varying observations. Exemplarily, the sequence of observations summarized in Table 6 is considered. The analysis starts with the a-priori case (a) that corresponds to the information available during the design phase. After construction of the infrastructure, measurements of the capacities of the structural elements are made, step (b). Thereafter, in the first two years of service, relatively low environmental loads are observed, together with the performance of the system elements, steps (c) and (d). In the third year, an extreme hazard event occurs. During the event, the only available information is that the hazard intensity is above a certain level, step (e). (For certain hazards, e.g., windstorms, predictions of $H$ could also be included prior to the event.) Immediately after the event, the available information is still incomplete and only the performances of two system elements are known, step (f). Finally, in the aftermath of the hazard event, the performance of the entire system and the exact hazard intensity become known, step (g).

Table 6. Sequence of observations in the infrastructure system example.

| Additional observations | Evidence entered in the rBN | |
|---|---|---|
| (a) | No observations (a-priori model) | No observation |
| (b) | Measurements of capacity | $M_{j4}$=180kN, $j = 1,2,3$ |
| (c) $i$=1,2,3,11,12,13. | Performance and hazard first year | $H(1)$=40kN; $E_i(1)$=1, |
| (d) $E_i(2)$=1, $i$=1,2,3,11,12,13. | Performance and hazard second year | $H(2)$=35kN; |
| (e) | Hazard third year | $H(3)$>70kN. |
| (f) | Performance of system elements | $E_3(3)$=1; $E_{13}(3)$=1. |
| (g) Performance and hazard third year | $H(3)$=85kN; $E_i(3)$=1, $i$=1,2,3,11,12,13. | |

The updated system reliability for all observation steps is presented in Figure 15 for a 10 year life span. Also, as a representative example, the updated reliability of structure 1 is presented in Figure 16. The values shown here are the reliability indexes corresponding to the probability of failure in a given year, consistent with the definition of $E_j(t)$.

At step (a), the unconditional system reliability slightly decreases with time due to the deterioration of the structural systems, which leads to a decreased capacity with respect to extreme environmental loads. At step (b), inclusion of the favorable information gained from the measurement of the capacities of the structural elements significantly increases the reliability of the infrastructure as well as each of the structures. The measurement results considered here are above the expected value of the capacities, but the difference is less than the combined standard deviation of the capacities and the measurement uncertainty. The observation of good performance in the first year (step c) slightly increases the reliability, due to a proof-loading effect with respect to the beam mechanism of the structural systems, as seen earlier in the life-cycle example. Because the observed environmental load value is low, there is no proof-loading effect related to the failure mechanisms involving the environmental loads. This is confirmed by the fact that inclusion of the information in step (d) has virtually no effect on the estimated reliabilities.

In step (e), only the immediate reliability index is of interest. The near-real-time availability of this assessment can provide crucial information to the emergency response personnel. This and step (f) demonstrate that any information becoming available during and in the immediate aftermath of a hazard event can be accounted for by the model. In the latter step, observation of survival of two system components leads to an increased reliability of the system. Finally, step (g) illustrates that after the event, and given that the entire infrastructure performed well, the reliability index is close to the one prior to the event. The small reduction in reliability is due to the model uncertainty on the environmental loading, represented by $U_H$, which causes dependence among the hazard events in different years. Observing a large value for $H(3)$ increases the probability of large hazards in future years.

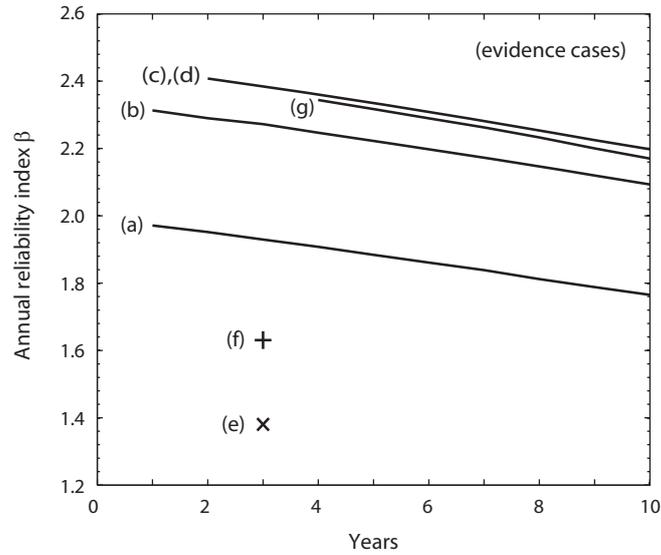

*Figure 15. Reliability index of the infrastructure system conditional on the sequence of observations.*

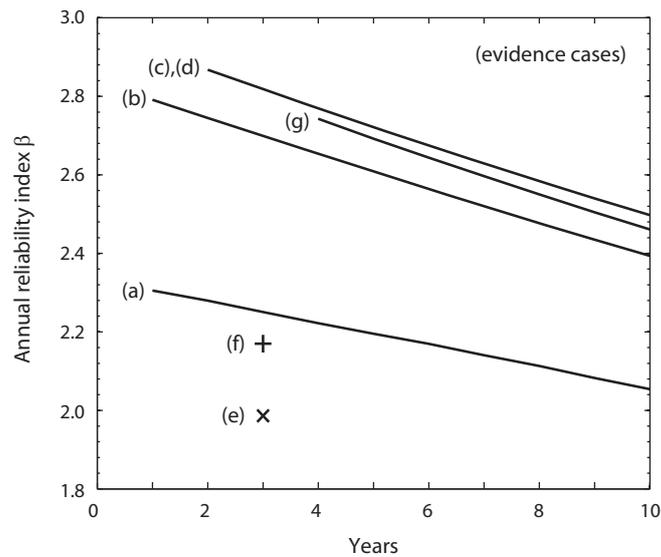

*Figure 16. Reliability of structure 1 conditional on the sequence of observations.*

The rBN enables the updating of all variables in the system, and thus facilitates learning about the model. To demonstrate this, we present in Figure 17 the posterior PDF of $U_H$, the variable representing the uncertainty in the model of the environmental load, for the sequence of relevant observations. The difference between the PDFs for steps (d) and (g) reflects the effect of observing a large hazard in year 3.

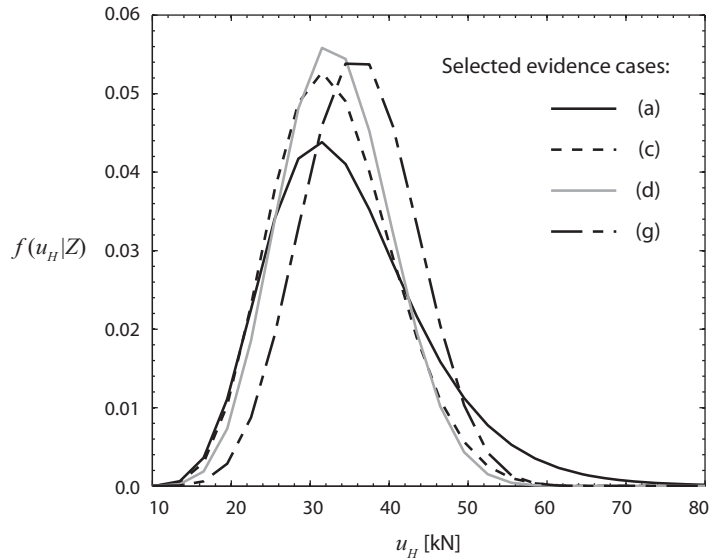

*Figure 17. The PDF of $U_H$ conditional on sequence of observations.*

## 4 Discussion

The presented examples illustrate the capabilities and potential of the eBN approach for reliability analysis of structures and infrastructure systems. From the viewpoint of application to decision support and optimization in near-real-time, where information becomes available gradually over time, a particular benefit is the possibility to apply exact inference algorithms for the analysis of the rBN. These algorithms are computationally efficient and robust, as long as the resulting rBN is not too complex. It is important to be aware of this limitation in the application and further development of the methodology.

As pointed out earlier, the sizes of the Markov envelopes in the eBN must be restricted. In the presented examples, this objective was achieved through the strategies suggested in the companion paper, in particular by selectively discretizing some of the continuous random variables and by introducing additional variables to reduce the number of parents to the discrete variables. The presented models can be further extended and larger (and more realistic) structural models and infrastructure systems can be considered. However, it is noted that certain extensions of the model require further development before being computationally feasible. For example, in the presented applications, the structural systems were modeled in detail at time zero, but were represented by single variables $Q_j(t)$ at later times. This choice was made to reduce the sizes of the Markov envelopes, but it does not allow detailed modeling of the deterioration of individual structural elements, and it does not enable direct inclusion of measurements and inspections of the elements of the structural systems at times $t > 0$. Additionally, when a structural system is subjected to different types of hazards, then $Q_j(t)$ must represent the capacity of the structure with respect to the joint effect of these hazards. Further work is required to determine an efficient representation in this case.

Furthermore, the computational complexity of the resulting rBN evaluations must be contained. Because the evaluations require the computation of the joint probability of the evidence (Straub and Der Kiureghian, 2010), the complexity depends on the evidence. All cases presented in this paper were computed efficiently, but, for the infrastructure analysis, combinations of evidence can be conceived for which the evaluation becomes too demanding for exact inference algorithms. This is the case, for example, when the performances of many infrastructure system elements at different times are observed without observing the related joint hazard. Although this case is not typical, strategies for dealing with such situations should be developed in the future. A potential solution is to switch to approximate inference algorithms in such cases, or use combinations of approximate and exact inference algorithms.

Finally, it is pointed out that many documented applications of structural reliability methods make use of ideas and concepts incorporated in the proposed eBN framework. In particular, for simple reliability analysis, the modeling underlying the eBN is quite intuitive. In these cases, the benefit of the eBN is that it represents a rigorous framework that supports the establishment, documentation and quality control of the models. Additionally, and this is arguably its most important benefit, the eBN facilitates the extension of simple structural reliability models to models with complex dependence structures and to decision graphs that support the optimization of mitigation actions under evolving information.

# 5 Conclusions

In this paper, the eBN methodology developed in the companion paper is successfully applied to time-varying reliability analysis of a single structural system and an infrastructure system composed of structural and nonstructural elements. The first example demonstrates how information obtained during the life-cycle of the structure from measurements of element capacities and from observations of past and current hazard events can be used to make optimal decisions regarding repair or replacement of the structure. The second example demonstrates how information gained during the life of an entire infrastructure system can be used to probabilistically update our state of knowledge about the future performance of the infrastructure and its elements. These calculations are performed through Bayesian updating in a computationally robust and efficient manner. The examples also illustrate the capabilities of the eBN to represent and account for various degrees of statistical and logical dependences in the models.

The presented methodology opens a new direction for research and development by combining tools from two well-developed methodologies: structural reliability methods, and BN combined with decision graphs. The methodology combines the computational power of SRMs for continuous variables, and powerful inference algorithms of BN for discrete variables. However, as discussed in the preceding section, additional work is needed to overcome both modeling and computational hurdles in order to solve more general and realistic structural and infrastructure systems under varying degrees of information. We hope this work will spark interest for additional research in this direction.

# 6 Acknowledgement

Parts of this work were supported by the Swiss National Science Foundation (SNF) through grant PA002-111428. Additional support from the Taisei Chair Fund at the University of California, Berkeley is acknowledged.